# The effect of heating rate and soaking time on microstructure of an advanced high strength steel

M.A. Valdes-Tabernero[1*], C. Celada-Casero[2], I. Sabirov[1], A. Kumar[2], R. H. Petrov[2, 3]

[1] IMDEA Materials Institute, Calle Eric Kandel 2, Getafe 28906, Madrid, Spain

[2] Department of Materials Science and Engineering, Delft University of Technology, Mekelweg 2, 2628 CD Delft, The Netherlands

[3] Department of Electrical Energy, Metals, Mechanical constructions & Systems, Ghent University, Technologiepark 46, 9052 Ghent, Belgium

*Abstract*

This work focuses on the effect of soaking time on the microstructure during ultrafast heat treatment of a 50% cold rolled low carbon steel with initial ferritic-pearlitic microstructure. Dilatometry analysis was used to estimate the effect of heating rate on the phase transformation temperatures and to select an appropriate inter-critical temperature for final heat treatments. A thorough qualitative and quantitative microstructural characterization of the heat treated samples is performed using a wide range of characterization techniques. A complex multiphase, hierarchical microstructure consisting of ferritic matrix with embedded martensite and retained austenite is formed after all applied heat treatments. In turn, the ferritic matrix contains recrystallized and non-recrystallized grains. It is demonstrated that the ultrafast heating generally results in finer microstructure compared to the conventional heating independently on the soaking time. There is a significant effect of the soaking time on the volume fraction of martensite of the ultrafast heated material, while in the samples heated with conventional heating rate it remains relatively unchanged during soaking. Recrystallization, recovery and phase transformations occurring during soaking are discussed with respect to the applied heating rate.

## 1. Introduction

[*]Corresponding author: Miguel Angel Valdés Tabernero.
Postal address: IMDEA Materials Institute, Calle Eric Kandel 2, Getafe 28906, Madrid, Spain. Phone: +34 91 5493422. E-mail: miguelangel.valdes@imdea.org

Steels have been the most widely used materials all over the world and are likely to remain a key material of choice in construction and manufacturing. Steel manufacturing is a multistage process, where the heat treatment of (semi-)final product (in form of sheet, rod, wire) to a great extent determines its microstructure and, hence, its properties. The current approach for steel heat treatment is based on homogenization of microstructure at elevated temperatures (either at austenitic or intercritical temperatures) and cooling with controlled rate often followed by further treatment to form the required microstructure [1]. In 2011, Cola *et. al.* [2] proposed an idea to apply ultrafast heat treatment for manufacturing advanced high strength steels (AHSS) with microstructures as heterogeneous as those processed via conventional heat treatments. This treatment was initially referred to as 'flash processing' [2], and other terms such as 'ultrashort annealing' [3] and 'ultrafast heating' [4–7] are widely used for this process in the recent literature. Ultrafast heat treatment is based on heating the material with the heating rate in the range of 100 to 1000 ºC/s to an intercritical temperature, very short soaking at this temperature followed by quenching. The whole process lasts just a few seconds and, therefore, is characterized by significantly reduced energy consumption compared to the conventional heat treatments [8].

The current state of the art in the effect of ultrafast heat treatment on the microstructure and properties of steels can be summarized as follows. The final microstructure of the ultrafast heat treated steels is determined by three major heat treatment parameters: heating rate, peak temperature and soaking time. Ultrafast heating typically results in grain refinement in interstitial free (IF) [9] and low carbon steels [3–5,10,11], thus, leading to higher mechanical strength. Increasing heating rate shifts the recrystallization temperature to higher values than the one measured at conventional heating rates of 10-20°C/s. Recovery and recrystallization processes concurrently occur during ultrafast heating, and increasing the heating rate decreases the recrystallized fraction of ferrite for a given temperature [5–7,12–14]. The martensite volume fraction in the heat treated steel tends to increase with increasing peak temperature [15]. The initial microstructure strongly influences the properties of steels after ultrafast heat treatment [5]. Particularly, the steels with the initial ferritic-pearlitic microstructure showed lower strength and higher ductility compared to the steels with the initial ferritic-martensitic microstructure [5]. The pre-heating stage at temperatures of 300-400 ºC has minor effects on the microstructure evolution during ultrafast heating, though increase of pre-heating temperature results in lower volume fraction of austenite, and hence martensite upon quenching, due to cementite spheroidization [12].

Microstructure evolution in steels during ultrafast heating and short soaking at the peak temperature is a very complex phenomenon, as it involves simultaneously recovery, recrystallization, grain growth, phase transformations and diffusion of alloying elements with carbon playing the key role. In most of the basic studies, the isothermal soaking time was taken as short as possible, 0.1- 0.2 s [5,7,12,13]. Such short soaking times cannot be reached during UFH processing of steel on the existing industrial lines and this is a significant obstacle for implementation of the ultrafast heating in steel industry. It was reported that longer isothermal soaking time (30 s) can erase the positive grain refining effect of the ultrafast heating [16]. However, in the current literature there are no systematic studies on the effect of the isothermal soaking time at the peak temperature on the microstructure and properties of steel after ultrafast heating. Fundamental understanding of microstructure evolution is required to enable an easy determination of the optimum soaking parameters for microstructural design in the ultrafast heat treated steels. Therefore, the main objective of the present work is to thoroughly study the effect of soaking time on the microstructure evolution during ultrafast heating of a low carbon steel. Conventional heating of the steel followed by detailed microstructural characterization is also performed for comparison.

## 2. Material and experimental procedures

### 2.1. *Material*

A low carbon steel with chemical composition of 0.19 % C, 1.61 % Mn, 1.06 % Al, 0.5 % Si (in wt. %) was selected for this investigation. Alloys with this composition are typically used in the automotive sector as transformation induced plasticity (TRIP) assisted steels, which belong to the 1$^{st}$ generation of AHSS [17–19]. Two kinds of heating experiments were performed: a) dilatometry measurements to determine phase transformation temperatures, and b) annealing tests to the intercritical temperature with varying soaking time followed by quenching. Both types of experiments are described in detail below.

### 2.2. *Dilatometry experiments*

As increasing heating rate shifts the recrystallization temperature to the higher values than the equilibrium one or the one measured at conventional heating rates [5,13]. Dilatometry measurements were carried out to determine the phase transformation temperatures $A_{C1}$ and $A_{C3}$ of the studied steel as a function of heating rate. For these experiments, specimens with dimensions of 10x5x1 mm$^3$ were machined from the as-received material. Tests were carried out in a Bähr DIL805A/D dilatometer (Bähr-Thermoanalyse GmbH, Hüll-Horst, Germany). Specimens were heated up to 1100 ºC with different heating rates (1, 10, 50 and 200 ºC/s) and holding time equal to 0.2 s. Heating rates above 200 ºC/s were not applied due to instability of the system in that range of heating rates. A *K*-type thermocouple was welded to the midsection of each specimen to measure their temperature during experiment. The material was then cooled down to room temperature at -300 ºC/s. The sample expansion/contraction during heating/cooling was recorded, and the obtained dilatometry curves were analyzed. The tangent intersection method was applied to determine the start ($A_{C1}$) and finish ($A_{C3}$) temperatures of austenite formation.

## 2.3. *Intercritical heat treatments*

For the intercritical heat treatments, strips of 100 mm in length and 10 mm in width were machined along the rolling direction and heat treated in a thermo-mechanical simulator Gleeble 3800. A *K*-type thermocouple was spot-welded to the midsection of each specimen. Two different types of heat treatment were applied. In both types, the thermal cycle was divided into five stages. On the first and second stages, the specimens were heated at 10 ºC/s to 300 ºC, followed by a soaking period of 30 s at 300 ºC. These stages simulate a preheating in some industrial continuous annealing lines to reduce the thermal stresses during heating. The third stage is heating from 300 ºC to the peak temperature of 860 ºC at two different heating rates, 10 ºC/s (conventional heating or CH) and 800 ºC/s (ultra-fast heating or UFH) followed by soaking at 860 ºC for 0.2 s. The processed specimens will be referred to as CH10-0.2s and UFH800-0.2s, respectively. Such a short soaking time (0.2 s) allows to eliminate the effect of annealing time on the microstructure and to focus entirely on the effect of heating rate. The last stage was to cool down the material to room temperature at ~160 ºC/s. The peak temperature of 860 ºC for intercritical annealing was selected based on the outcomes of the dilatometry measurements (see Section 3.1).

To study the effect of soaking time at both heating rates (CH and UFH), additional heat treatments were performed with higher soaking time (1.5 s and 30 s). The new generated conditions are referred to as CH10-1.5s and CH10-30s for the CH treatment, and UFH800-1.5s and UFH800-30s for the UFH treatment. All applied thermal cycles are schematically presented in (**Figure 1**). In all samples, a minimum length of 10 mm of the homogeneously heat treated zone was verified by microhardness measurements.

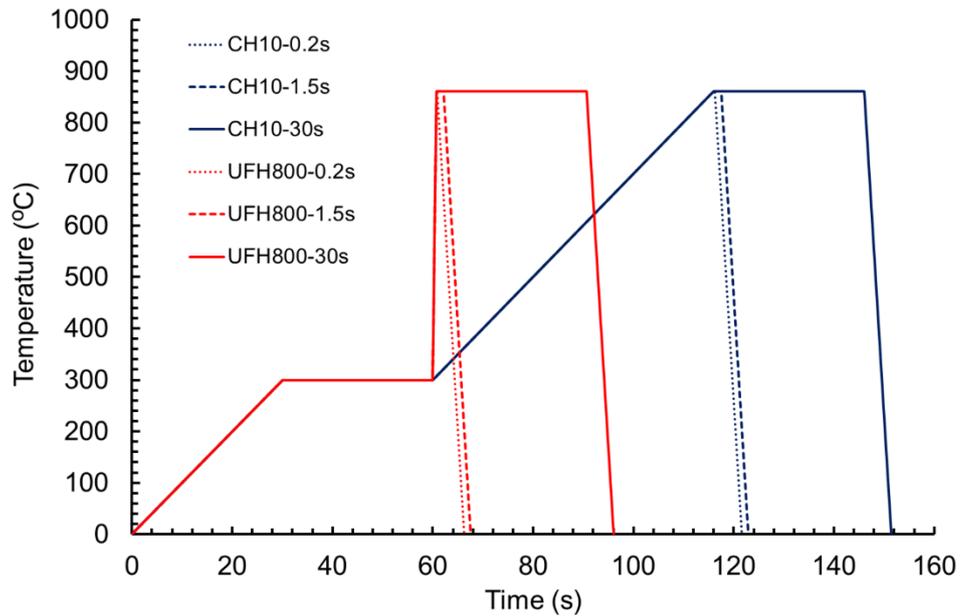

**Figure 1**: Schematic representation of the different heat treatments applied to the studied material. (For interpretation of the references to color in this figure, the reader is referred to the web version of this article).

### 2.4. *Microstructural characterization*

A thorough microstructural characterization of the samples heat treated in a thermo-mechanical simulator (**Figure 1**) was performed. Specimens for scanning electron microscopy (SEM) studies were ground and polished to a mirror-like surface applying standard metallographic techniques with final polishing using OP-U (colloidal silica). The polished specimens were etched with 3 vol.% Nital solution for 10 s. Examination of the microstructure was performed using a FEI Quanta™ 450 FEG-SEM operating at an accelerating voltage of 15 kV. Microstructure was observed on the RD–ND plane.

Specimens for electron backscatter diffraction (EBSD) analysis were ground and polished following the same procedure as for SEM images. Orientation imaging microscopy (OIM) studies were performed using a FEI Quanta™ Helios NanoLab 600i equipped with a NordlysNano detector controlled by the AZtec Oxford Instruments Nanoanalysis (version 2.4) software. The data were acquired at an accelerating voltage of 18 kV, a working distance of 8 mm, a tilt angle of 70º, and a step size of 65 nm in a hexagonal scan grid. The orientation data were post-processed using HKL Post-processing Oxford Instruments Nanotechnology (version 5.1©) software and TSL Data analysis version 7.3 software. Grains were defined as a minimum of 4 pixels with a misorientation higher than 5º. Grain boundaries having a misorientation ≥ 15º were defined as high-angle grain boundaries (HAGBs), whereas low-angle grain boundaries (LAGBs) had a misorientation < 15º. Textures are represented as orientation distribution functions (ODFs) using Bunge notation [20]. The ODFs were derived from the EBSD scans by superimposing Gaussian distributions with a half-width of 5°. The resulting ODF was represented as a series expansion of spherical harmonics functions with a maximum rank of the expansion coefficient L = 16. Texture and grain size calculations were made using scans having area of ~ 6000 µm$^2$ which contains at least 1100 grains. The volume fractions of transformed/untransformed grains and recrystallized/recovered ferritic grains were determined by a two-step partitioning procedure described in [5,21]. In this procedure, grains with high (> 70º) and low (≤ 70º) grain average image qualities are separated in a first step, allowing to distinguish between untransformed (ferrite) and transformed (martensite) fractions, respectively. In the second step, recrystallized and non-recrystallized ferritic grains are separated using the grain orientation spread criterion: Grains with orientation spread below 1º are defined as the recrystallized grains, while grains with an orientation spread above 1º are defined as the non-recrystallized ones [22]. It should be noted that another grain average misorientation based criterion was employed in our recent report [14] for separation of recrystallized/non-recrystallized grains. Comparison of these two different criteria via analysis of numerous EBSD scans carried out in this work has shown, that the criterion utilized in the present manuscript yields better results. The microstructure was characterized on the plane perpendicular to the sample transverse direction (the RD–ND plane).

X-ray diffraction (XRD) experiments were carried out to determine the retained austenite volume fraction and its carbon concentration. Specimens with a surface of 10 x 5 mm$^2$ were prepared following the same procedure as for the EBSD analysis. The measurements were performed using a Bruker D8 Advance diffractometer (Bruker AXS, Karlsruhe, Germany) equipped with a VANTEC position sensitive detector and using Co K$_\alpha$ radiation ($\lambda$ = 1.78897 Å), an acceleration voltage of 45 kV and current of 35 mA. The measurements were performed in the 2θ range from 45º to 130º with a step size of 0.035º and a counting time per step of 3 s. The volume fraction of retained austenite was calculated using the Jatczak model as described in [23]. The austenite carbon concentration, $X_c$, was estimated from its lattice parameter, $a_\gamma$. The latter was determined from the austenite peak position as [24]:

$$a_\gamma = 0.3556 + 0.00453\, X_c + 0.000095\, X_{Mn} + 0.00056\, X_{Al} \qquad (1)$$

where $a_\gamma$ is the austenite lattice parameter in nm and $X_i$ represents the concentration of the alloying element *i* in wt. %. The effect of silicon and phosphorous is not taken into account, as it is negligible compared to other elements considered in Eq. (1).

In order to carry out a thorough characterization of nanoscale constituents in a rapid manner, in 2012 Keller *et al.* proposed a novel approach called transmission Kikuchi diffraction (TKD) analysis [25]. It is based on performing an EBSD analysis in transmission mode. The method requires very thin samples, similar to those for TEM characterization, and a conventional SEM equipped with EBSD detector. It can also be combined with transmission electron microscopy (TEM) analysis. Due to the low thickness of sample, typical SEM voltages are sufficient for electrons to interact with the material and pass through, to finally be captured by the EBSD detector. TKD offers better spatial resolution (< 10 nm) than EBSD, allowing the resolution of nanoscale microstructural constituents having 10-30 nm in size [26,27]. It has been successfully used to analyze oxides and nitrides in aluminium alloys [28] and stainless steels [29,30], as well as martensite and retained austenite in bainitic steels [31]. In this work, for TKD and TEM studies, the samples were ground to a thickness of 100 µm and disks of 3 mm in diameter were subsequently punched out. The disks were further thinned in a Struers Tenupol-5 via twin-jet electropolishing until a central hole appeared. The used electrolyte was composed of 4 % vol. $HClO_4$ in 63 % water-diluted $CH_3COOH$ under 21 V at 20 ºC and a flow rate equal to 17. TKD data were collected by an EDAX-TSL EBSD system attached to a FEI Quanta™ 450-FEG-SEM under the following conditions: accelerating voltage of 30 kV, working distance of 4 mm, tilt angle of - 40°, a beam current of 2.3 nA corresponding to the FEI spot size of 5, aperture size of 30 μm. TKD measurements were performed with the step size of 10 nm. The orientation data were post-processed using TSL Data analysis version 7.3 software. TEM images were acquired in a Jeol (S)TEN JEM-2200FS operated at 200 kV and equipped with an aberration corrector of the objective lens (CETCOR, CEOS GmbH) and a column electron energy filter (omega type). XRD, TEM and TKD measurements were performed on samples CH10-0.2s, UFH800-0.2s, UFH800-1.5s and UFH800-30s.

## 3. Results and discussion

*3.1. Dilatometry*

**Figure 2**a represents the typical dilatometry curves for the samples tested with different heating rates. The $A_{C1}$ temperature was determined at 5 % volume fraction of the transformed phase calculated by the lever rule (as shown in **Figure 2**b). Such relatively high percentage of the transformed phase was selected as a criterion due to complexity of the microstructure evolution during heating, which involves various processes (carbide dissolution, recovery and recrystallization of ferrite, formation of austenite as observed in [32–34] and described in Section 3) resulting in $A_{C1}$ temperature range. Once the sample is fully austenitic at the $A_{C3}$ phase transformation temperature, the expansion becomes linear with the temperature. The martensite start temperature $M_S$ corresponds to the point on the dilatation curve, where the contraction of austenite during quenching is replaced by expansion due to the formation of martensite. As it is seen from **Table 1**, all three transformation temperatures, $A_{C1}$, $A_{C3}$ and $M_S$, tend to increase with the increasing heating rate.

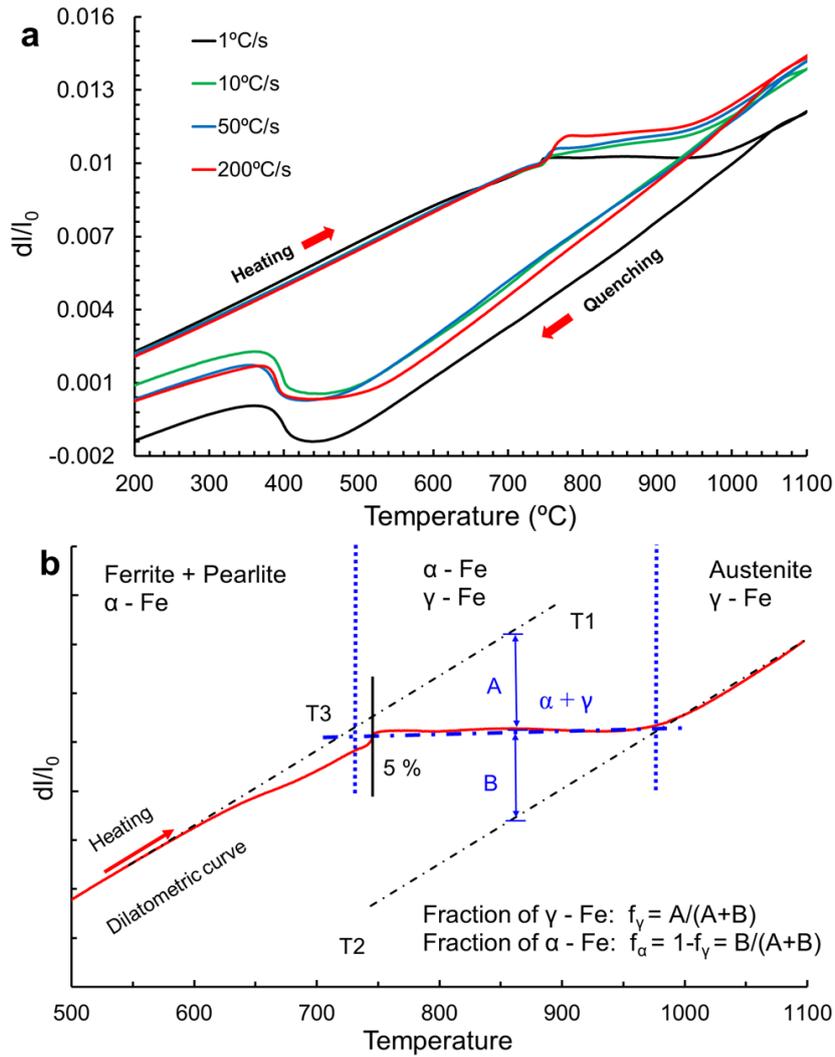

**Figure 2**: a) Dilatometry curves from dilatometry tests with different heating rates; b) Schematic diagram of an experimental dilatometry curve (measured at 1 °C/s) to calculate $A_{C1}$ and $A_{C3}$ temperatures via tangent intersection principle and lever rule. (For interpretation of the references to color in this figure, the reader is referred to the web version of this article).

**Table 1**: Effect of the heating rate on the phase transformation temperatures: $A_{C1}$, $A_{C3}$ and $M_S$.

| Heating rate (ºC/s) | $A_{C1}$ (ºC) | $A_{C3}$ (ºC) | $M_S$ (ºC) |
|---|---|---|---|
| 1   | 738 | 968 | 483 |
| 10  | 760 | 969 | 489 |
| 50  | 781 | 971 | 498 |
| 200 | 793 | 983 | 530 |

For the $A_{C1}$, the pronounced increase from 738 to 781 °C occurs at the lower heating rates ranging from 1 °C/s to 50 °C/s. On the other hand, the $A_{C3}$ temperature just slightly grows from 968 to 971 °C in that temperature range jumping up to 983 °C at 200 °C/s. It can be hypothesized, that this variation of the $A_{C1}$ temperature is determined mainly by nucleation and growth rate of austenitic grains. The nucleation rate at the given elevated temperature grows with the increasing heating rate, since the latter suppresses the recovery effects, resulting in higher density of lattice defects at the given temperature, which, in turn, promote phase nucleation. The growth rate of the nucleated austenitic grains is controlled by carbon diffusion [7] and solute drag effect (by Mn atoms in the studied steel) [35]. Therefore, at the early stages of phase transformation, the austenite volume fraction at the given temperature decreases with increasing heating rate. Both factors result in increasing $A_{C1}$ temperature with rising heating rate. It should be noted that similar results were earlier published in [36]. In this study, a linear dependency of $A_{C1}$ on the heating rate (**Figure 3**) on the semi-log plot is observed. Similar tendency of $A_{C1}$ on the heating rate for ferritic-pearlitic microstructure has been reported in [37,38]. The nucleation and growth depend on the heating rate exponentially [38]. Moreover, the extrapolation of this behavior to low heating rates (0.2 °C/s) shows an equilibrium temperature of 720 °C, which is very close to the theoretical one (723 °C), thus confirming the linear character of this dependence. Therefore, this approach can also be used to predict the $A_{C1}$ temperature at high heating rates. Particularly, for 800 °C/s, the $A_{C1}$ temperature is about 808 °C (**Figure 3**). On the other hand, the dependence of $A_{C3}$ temperature on the heating rate is less pronounced. Similar observations were reported earlier in [39]. Therefore, the intercritical temperature of 860 °C was selected as the peak temperature for both CH and UFH treatments (see Section 2.3).

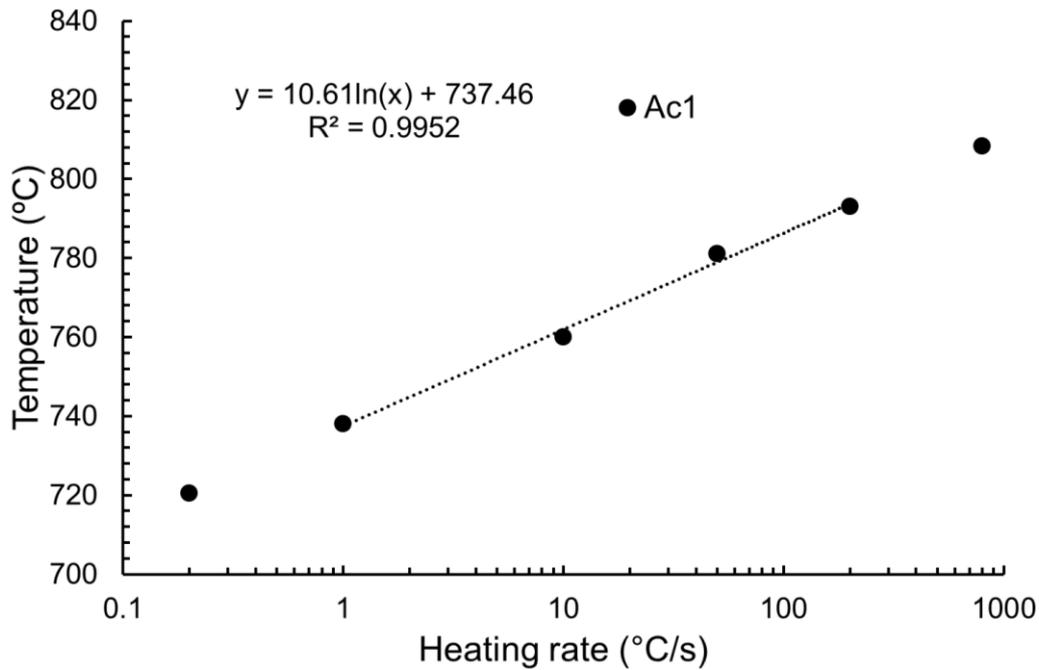

**Figure 3**: Effect of heating rate on the $A_{C1}$ temperature.

Increasing heating rate during heat treatment with full austenitization followed by immediate cooling leads to increment of the $M_S$ temperature. This effect is produced because the higher applied heating rate results in the higher amount of defects in the microstructure induced by cold rolling. As recovery is diffusion controlled [40], higher density of lattice defects is retained in the microstructure due to shorter time at elevated temperatures. This effect was observed previously in [41,42]. In addition, at high heating rates carbides remain undissolved in the microstructure, leading to a formation of austenite with lower carbon content and, hence, a higher $M_S$ compared to the conventional heating rates. Therefore, the steepest increment on $M_S$ is produced, when heating rate grows from 50 ºC/s to 200 ºC/s leading to an increase of transformation temperature from 498 ºC to 530 ºC. On the other hand, in the range of lower heating rates from 1 to 50 ºC/s the $M_S$ temperature just slightly varies.

*3.2. SEM characterization*

The supplied material shows a typical cold rolled microstructure consisting of elongated grains of deformed ferrite with volume fraction of 76 % and pearlite with volume fraction of 24% (**Figure 4**).

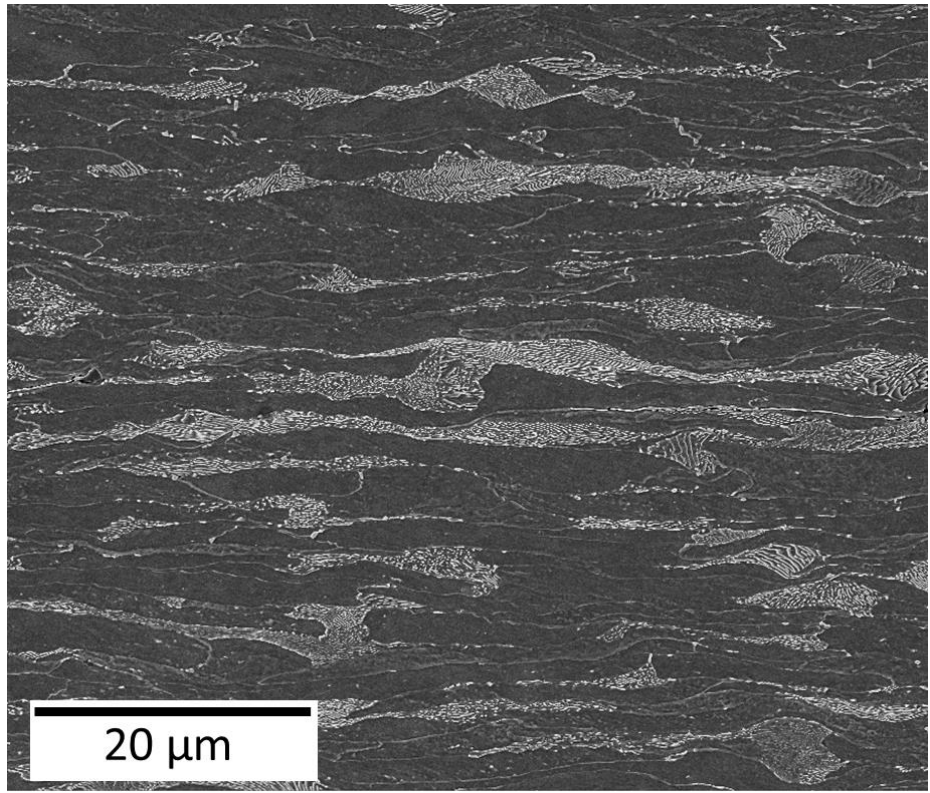

**Figure 4:** Initial ferritic-pearlitic microstructure of the steel after 50 % cold reduction, being ferrite in grey and pearlitic colonies in white.

The microstructure after CH treatment with soaking time of 0.2 s, 1.5 s and 30 s is presented in **Figure 5**a, b, c, respectively, whereas remaining images illustrate the microstructure after UFH treatment. In all cases, the material presents a complex microstructure formed by a ferritic matrix (consisting of recrystallized and recovered ferritic grains) with embedded martensite and retained austenite grains. However, it strongly depends on the applied heat treatment parameters. During CH treatment, the material presents a similar microstructure independently on the soaking time, while the latter has very significant effect on the microstructure formed after UFH treatment.

CH treatment generates a ferritic matrix with homogeneous microstructure consisting of equiaxed grains, as previously observed in [5]. On the other hand, UFH results in the matrix microstructure consisting of fine equiaxed grains and larger elongated grains surrounded by martensitic grains. The large grains may grow from the heavily deformed ferrite located in the vicinity of pearlite colonies, as the latter are not able to accumulate high plastic strain during rolling. Hence, the higher energy stored in the heavily deformed ferritic areas leads to a faster grain growth [40]. Some Widmanstätten ferritic grains are also observed in the UFH samples after soaking for 1.5 and 30 s (marked by white arrows on **Figure 5**h, i) possibly formed at the early stages of cooling. Those ferrite plates are surrounded by bainite.

Spheroidized cementite (SC) is also observed in samples UFH-0.2s and UFH-1.5s (marked by red dashed arrows on SEM micrographs presented on **Figure 5**). It is related to the short time (0.2 – 1.5 s) of the heat treatment, as reported previously by Castro Cerdá *et al*. [5,43], and fully dissolved after soaking for 30 s. A very small region with spheroidized cementite particles was also observed in the CH-0.2s sample, although its amount is negligible (**Figure 5**a).

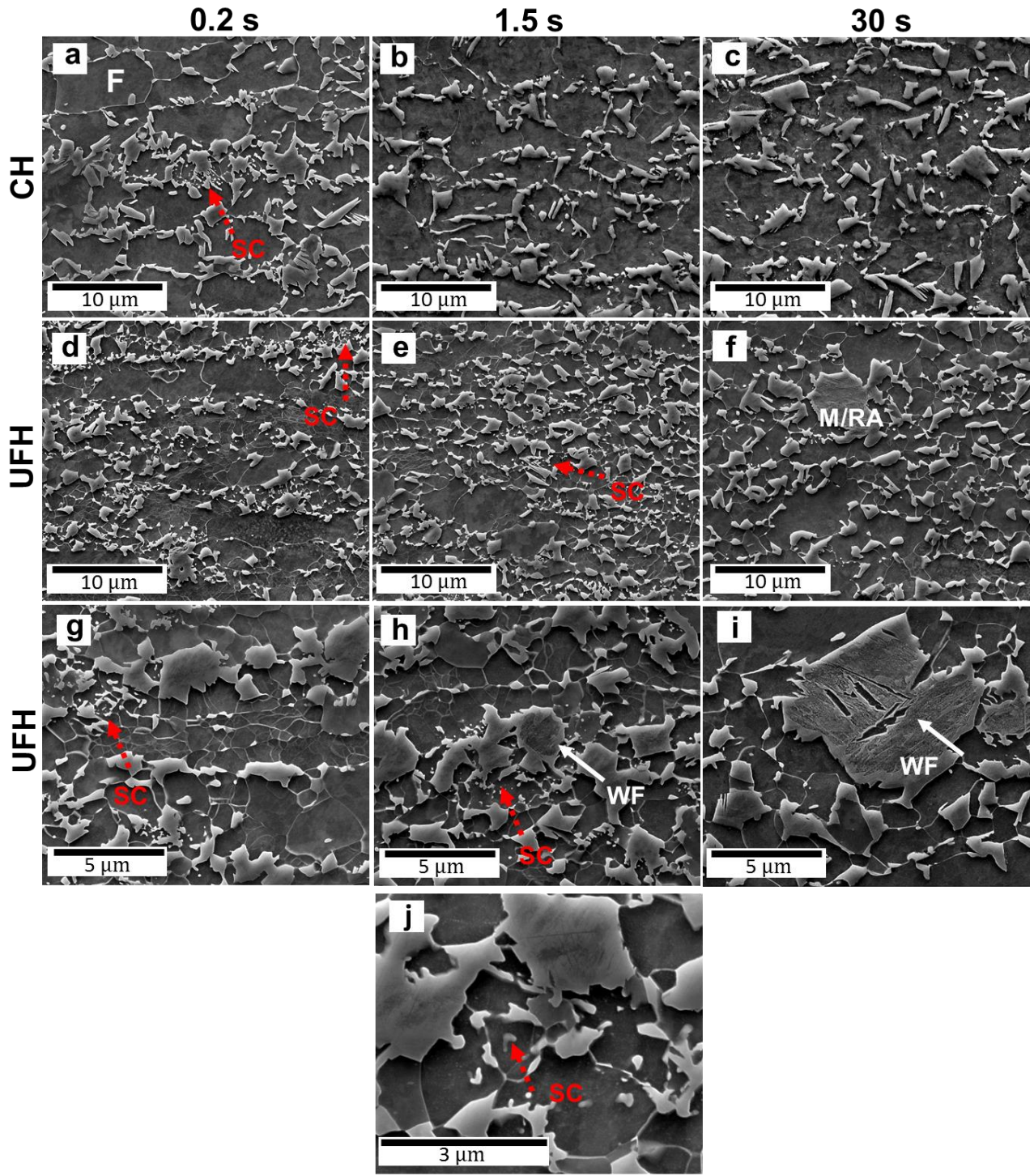

**Figure 5**: SEM micrographs showing the effect of heating rate (10 and 800 ºC/s) and soaking time (0.2 to 30 s) on the microstructure: a), b) and c) correspond to 10 ºC/s for 0.2, 1.5 and 30 s, respectively; d), e) and f) correspond to 800 ºC/s for 0.2, 1.5 and 30 s, respectively. Higher magnification images g), h) and i) show microstructures heated at 800 ºC/s for 0.2, 1.5 and 30 s, respectively; j) higher magnification image of spheroidized cementite (SC) in the sample heated at 800 ºC/s for 1.5 s. Spheroidized cementite is marked by dashed red arrows, while white arrows

indicate Widmanstätten ferrite (WF). Ferrite is marked as F, and M/RA stands for martensite/retained austenite. Etched with Nital (3%).

*3.3. EBSD characterization*

EBSD technique was used to precisely quantify and characterize the different microconstituents formed in the material after both heat treatments. The results of EBSD analysis are outlined in **Table 2**. CH treatment leads to a microstructure mainly formed by a ferritic matrix, whose volume fraction remains constant (~ 86–87 %) and martensite volume fraction slightly increases from 10.6 % to 12.5 % with the soaking time. As volume fraction of ferrite does not vary with soaking time (i.e. the amount of intercritical austenite formed at the peak temperature does not depend on the soaking time), the martensite increment can be attributed to the partial transformation of austenite into martensite by deformation during sample preparation. This indicates that retained austenite is less stable caused by the homogenization of carbon distribution in its interior after longer soaking times. Although the UFH process generates similar microstructure with the same microstructural constituents, there are significant variations in the volume fractions of different phases with respect to the CH treatment. The volume fraction of ferrite noticeably decreases with increasing soaking time from 90.9 % at 0.2 s to 75.9 % at 30 s, while the volume fraction of martensite shows the opposite trend. As the volume fraction of retained austenite remains stable (2.1 – 2.2 %), it is possible to assure that the decrease of ferrite fraction is directly associated to the formation of martensite. On the other hand, the difference in ferrite and martensite volume fractions between CH and UFH conditions can be explained by the spheroidization of cementite during heating. First, the nucleation of austenite occurs at the α/cementite interface [44]. With conventional heating (CH), the cementite spheroidizes [7] reducing the amount of preferable sites for austenite formation and resulting in longer soaking time to reach the equilibrium. The main fraction of the inter-critical austenite is transformed into martensite during cooling. On the other hand, during UFH treatment the peak temperature is reached in less than 1 s which dramatically reduces the amount of spheroidized cementite and, thus, increases the driving force for austenite nucleation at the more favorable α/cementite interfaces.

Table 2: Effect of the heating rate and soaking time on the volume fractions of phases present in the studied material.

| Condition (s) | CH | | | UFH | | |
|---|---|---|---|---|---|---|
| | 0.2 | 1.5 | 30 | 0.2 | 1.5 | 30 |
| Ferrite (%) | 86.3 ± 2.4 | 87.4 ± 2.7 | 85.8 ± 1.6 | 90.9 ± 4.0 | 85.3 ± 2.8 | 75.9 ± 4.6 |
| Martensite (%) | 10.6 ± 1.7 | 10.8 ± 1.6 | 12.5 ± 1.6 | 6.9 ± 3.2 | 12.6 ± 3.1 | 22.0 ± 3.0 |
| Retained austenite (%) | 3.1 ± 0.7 | 1.8 ± 0.6 | 1.7 ± 0.1 | 2.2 ± 0.4 | 2.1 ± 0.3 | 2.1 ± 1.9 |

The morphology of the ferritic matrix in the CH and UFH heat treated samples also presents significant differences. The EBSD analysis revealed both recrystallized and recovered grains in the ferritic matrix. **Figure 6** represents the fraction of recrystallized ferrite in the ferritic matrix for all analyzed conditions. It is seen that, while the CH treatment leads to a homogeneous ferritic matrix, where almost 90 % of ferrite is recrystallized, the UFH processing generates a matrix microstructure formed by recrystallized and non-recrystallized (i.e. recovered) ferritic grains. After UFH treatment, the volume fraction of recrystallized ferrite increases from ~50 % after 0.2 s to ~67 % after 30 s. So, while the recrystallization process is completed during CH treatment already after soaking for 0.2 s, it is delayed during UFH process. Similar observations were previously reported in [43,45,46]. This effect is due to the competition of different processes, such as austenite formation and further grain growth, reducing the driving force for recrystallization. For short soaking time (0.2 s), the recrystallization is the controlling process, which results in a very low martensite volume fraction (**Table 2**), similar to the CH treatment, and a significant volume fraction of recrystallized ferrite present in the material (**Figure 6**). However, after soaking for longer time (1.5 – 30 s), other processes become dominant over recrystallization, such as the nucleation and growth of austenite into ferrite and ferrite grain growth [10,16]. The first effect results in the higher volume fraction of martensite present in the UFH800-30s (**Table 2**) and the decrease in volume fraction of recrystallized ferrite with increasing soaking time from 1.5 to 30 s (**Figure 6**). The latter effect is discussed more in detail below (**Figure 7** and **Figure 8**).

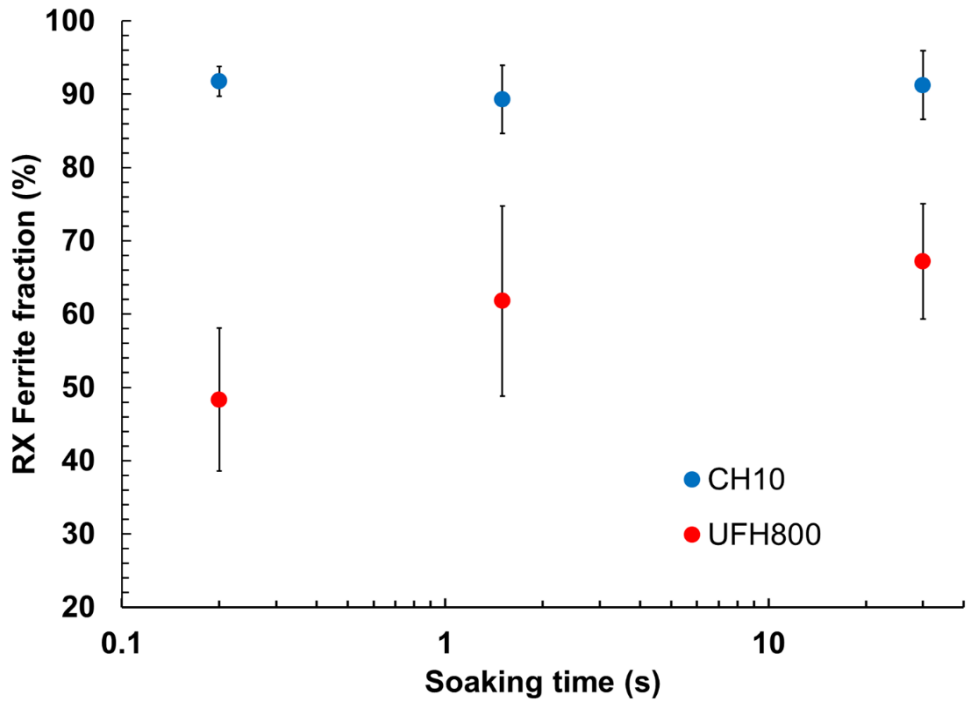

**Figure 6**: Evolution of volume fraction of recrystallized ferrite with respect to the total fraction of ferrite with heating rate and soaking time. (For interpretation of the references to color in this figure, the reader is referred to the web version of this article).

**Figure 7** represents the IPF maps for recrystallized (a, b, c) and non-recrystallized (d, e, f) ferrite after UFH for 0.2, 1.5 and 30 s, respectively. It is seen in **Figure 7**a, b, that the vast majority of the grains are in the early stage of growth, presenting a size $\leq 1.5$ μm, although it is possible to observe grains which have fully recrystallized and grown, i.e. grains without LAGBs and with low misorientations in their interior. This observation was also reported by Castro Cerda *et al.* [5]. When soaking time increases to 30 s, the fraction of fine grains decreases due to their growth, and the presence of larger grains is more evident (**Figure 7**c). The non-recrystallized grains demonstrate significant misorientation in the interior of the grains indicating formation of substructure independently on the applied soaking time (**Figure 7**d, e, f).

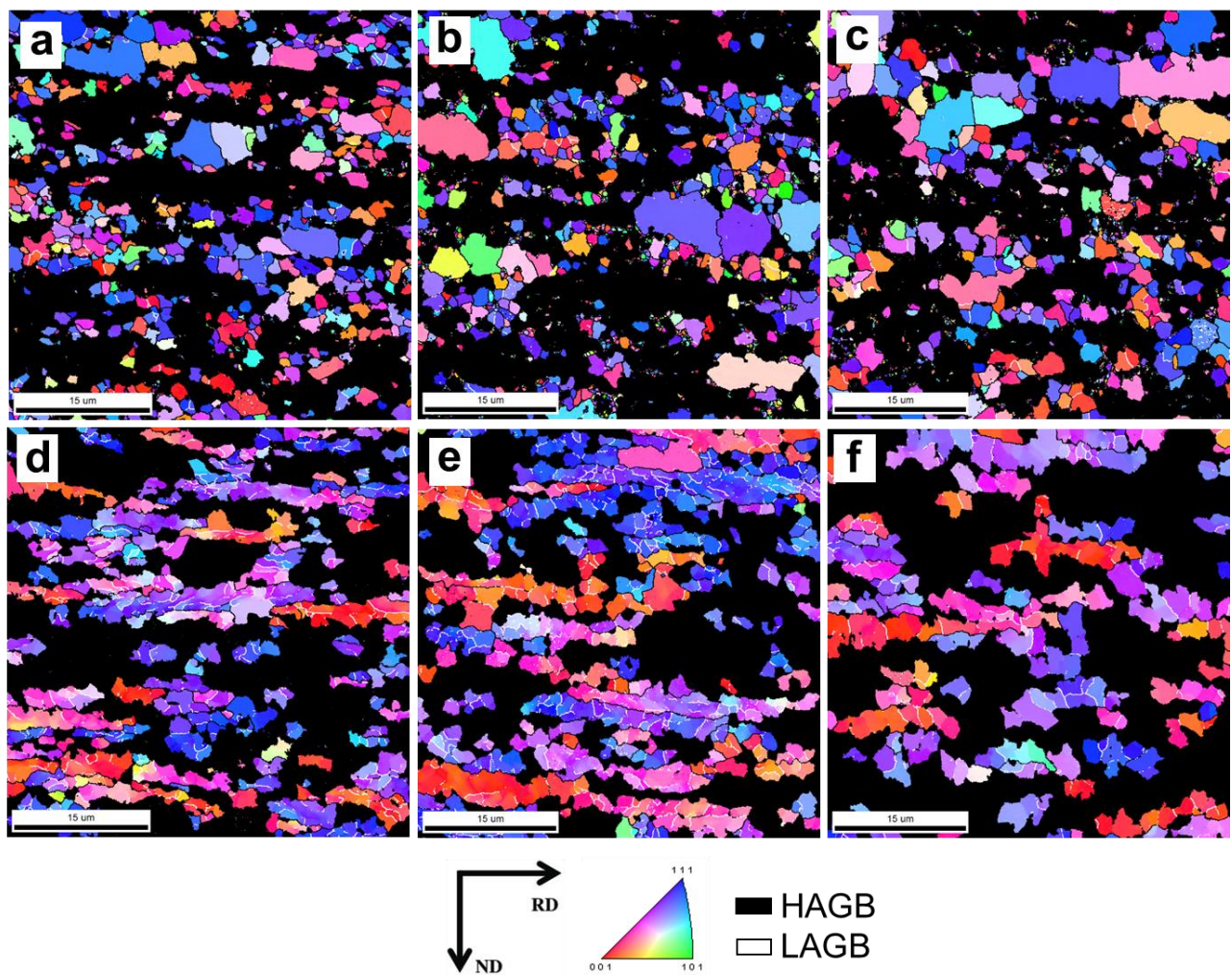

**Figure 7**: IPF maps after UFH treatment showing the recrystallized (a, b, c) and non-recrystallized (d, e, f) ferrite after 0.2, 1.5 and 30 s, respectively. HAGBs are shown in black and LAGBs in white. (For interpretation of the references to color in this figure, the reader is referred to the web version of this article).

The evolution of the grain size distribution for recrystallized ferrite is clearly visible and quantified in **Figure 8**a, b, c, where the grain size is plotted vs. the area fraction for the UFH800-0.2s, UFH800-1.5s and UFH800-30s, respectively (blue lines). It is observed that the mean peak shifts to higher values and widens. For instance, in the samples UFH800-0.2s and UFH800-1.5s the fraction of grains with a size below 1.5 μm is 52 % and 56 %, respectively, while after longer soaking it decreases to 36 % indicating the growth of the small grains nucleated at shorter times. A second peak at higher grain size is noticeable indicating the presence of the large grains mentioned above. The intensity of the second peak decreases with soaking time, as the microstructure becomes more homogeneous (**Figure 8** c). The histogram of grain size distribution for non-recrystallized ferritic grains (red lines in **Figure 8**) presents a similar character in comparison to the recrystallized ones. The primary peak shifts to the higher values becoming wider, when soaking time is increased. The fraction of grains having size above 2.5 µm increases from 59 % at 0.2 s to 68 % at 1.5 s to 73 % after 30 s. This effect can be produced by the coalescence of grains after partial recrystallization indicated by the presence of HAGBs. Nevertheless, the non-recrystallized grains are larger compared to the recrystallized ones after all soaking times. On the other hand, the ferritic matrix in the CH condition is formed mainly by recrystallized equiaxed grains, and its microstructure is not affected by soaking time (**Figure 8**d).

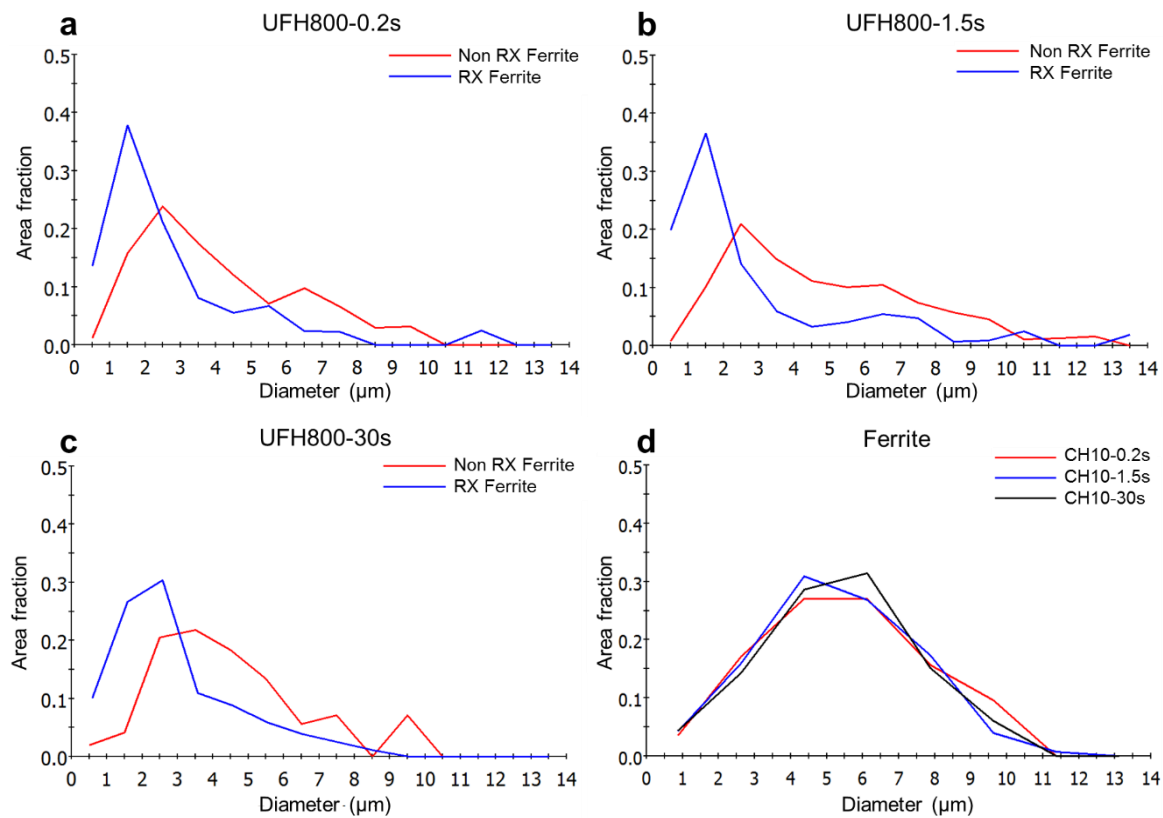

**Figure 8**: a), b), c) Representation of the equivalent circle diameter (ECD) versus area fraction for recrystallized (RX) and non-recrystallized (Non RX) ferrite after UFH with soaking for 0.2, 1.5 and 30 s, respectively; d) grain diameter versus area fraction for ferrite after CH treatment. Data are obtained from the EBSD measurements. (For interpretation of the references to color in this figure, the reader is referred to the web version of this article).

It is well known that high heating rates lead to a smaller grain size [6,10,13,47,48], as it is shown for the studied steel in **Figure 8**. This is caused, among other reasons, by the short time given to the α/α interface to grow. On the one hand, after CH treatment the recrystallization and grain growth processes are completed independently on the applied soaking time. The grain size is also not affected by soaking time, as intercritical austenitic grains act as barriers for the ferritic grains suppressing their further growth. On the other hand, the UFH treated conditions show a bimodal distribution of grain size. The presence of the two differentiated regions on the histograms can be rationalized by the interplay of two main effects:

(1) the effect of the initial heterogeneous microstructure related to different amounts of strain accommodated by individual ferritic grains, as shown in **Figure 4**;

(2) the effect of heating rate. A higher heating rate results in a recrystallization process taking place at higher temperatures, as discussed above, and, thus, in a higher nucleation rate due to the high density of defects [13,43,48].

The nuclei formed within the highly deformed areas possess higher driving force to grow and coalesce due to the high energy stored during cold rolling, resulting in the larger grains. On the other hand, nuclei generated within the less deformed regions present reduced driving force for growth. Moreover, due to the short time of the heat treatment, remains of individual cementite particles (which were not completely dissolved during inter-critical annealing) located at grain boundaries effectively pin grain boundaries suppressing grain growth and coalescence [49–51] (**Figure 5**g, h, i). As the material is heated up to an intercritical temperature, another important factor comes into play: Formation of austenite and its growth competes for the energy stored in the material. The austenitic grains nucleate in carbon enriched areas, i.e. within pearlitic colonies. It can be assumed that the intensive nucleation of austenitic grains takes place within pearlitic colonies which were severely deformed, rotated or broken during cold rolling, resulting in reduction of distance between cementite plates. As is well known, the austenite nucleation rate is inversely proportional to the inter-lamellar spacing of pearlite [12]. The austenite grows firstly into the pearlite until it is dissolved and then into ferrite, as it is seen in **Figure 5**. Competition of all these processes during UFH treatment results in the microstructure with finer grains (**Figure 5**, **Figure 8**).

**Figure 9** represents the equivalent circle diameter of martensite plotted versus area fraction. For the CH condition, at short soaking time (0.2 s) most of the martensite grains were formed from ultrafine austenitic grains, as the major peak lies below 1 μm (**Figure 9**a). Increasing soaking time up to 1.5 s, the curve shifts to the right, indicating the growth of the earlier formed nuclei. Finally, after annealing for 30 s, the decrease of the main peak intensity is accompanied by increase in the area fraction at 3 μm, displaying that the austenite has entered the growth stage after the nucleation after short soaking times. In the case of the UFH800-0.2s, the curve is similar to the CH condition with the same soaking time. However, the fraction of larger grains having a size of 4-5 µm increases. This behavior can indicate that the austenite nucleation is accompanied by a growth, due to the fact that the material has higher energy compared to the CH condition because of the low amount of spheroidized cementite and the higher carbon gradients present in the material, both produced by the rapid heating. It is more pronounced after 1.5 s, where the main peak has reduced, but there is an increase of the fraction of larger grains. The result of this effect is the rise of the martensite fraction in the overall microstructure. Finally, after 30 s the peak spreads to higher values, as it happens in the ferrite, showing an intense growth of the austenite grains during soaking.

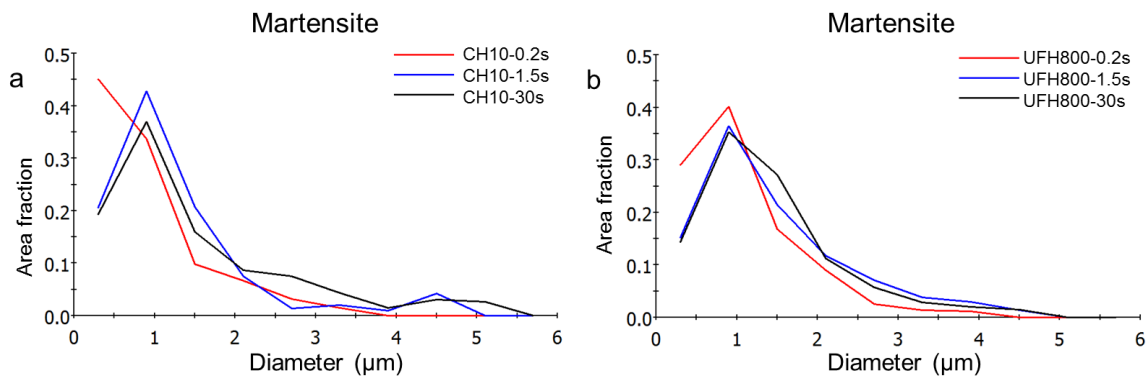

**Figure 9**: Martensite ECD vs area fraction for CH (a) and UFH (b) for different soaking times: 0.2 s, 1.5 s and 30 s.

*3.4. Texture analysis*

To analyze evolution of the preferable crystallographic orientation of ferritic grains, texture analysis was carried out for all studied conditions. **Figure 10**a represents the ideal positions of the most important texture components in BCC lattice, while **Figure 10**b shows the orientation distribution function (ODF) of the initial cold-rolled material. **Figure 10**c, d, e display the ODFs for the CH samples annealed for 0.2, 1.5 and 30 s, respectively, while **Figure 10**e, f, g represent the UFH conditions soaked for 0.2, 1.5 and 30 s, respectively. The initial cold-rolled material is represented by the ND {111}⟨uvw⟩ and RD {hkl}⟨110⟩ fibers, with a maxima corresponding to {111}⟨110⟩ components. Similar texture was found previously in cold-rolled low carbon steels [52,53]. On the other hand, the CH samples (**Figure 10**c, d, e) present an opposite curvature in the ND fiber compared to the initial cold-rolled microstructure and lower intensity in the RD fiber. Both effects can be associated with the recrystallization in the ferritic matrix [4]. In the UFH conditions (**Figure 10**f, g, h), the ODFs display texture similar to the initial cold-rolled condition (**Figure 10**b), with a strong intensity in the ND fiber components, indicating that complete recrystallization has been delayed. However, its intensity is reduced with increasing soaking time. This effect can be attributed to onset of recrystallization during intercritical annealing for > 1.5 s and increasing fraction of recrystallized grains with soaking time revealed by EBSD analysis (**Figure 6**, Section 3.3), as the initial ND fiber grains in the cold rolled steel present the higher stored energy [54].

The alpha fiber in the UFH treated material is also affected by soaking time. While a significant fraction of gamma fiber components recrystallized during UFH due to higher energy stored during cold rolling (compared to the alpha fiber components) [55,56], a lower fraction of alpha possesses energy (i.e. driving force) sufficient for recrystallization. So the RD fiber intensity is retained to large extent during UFH treatment.

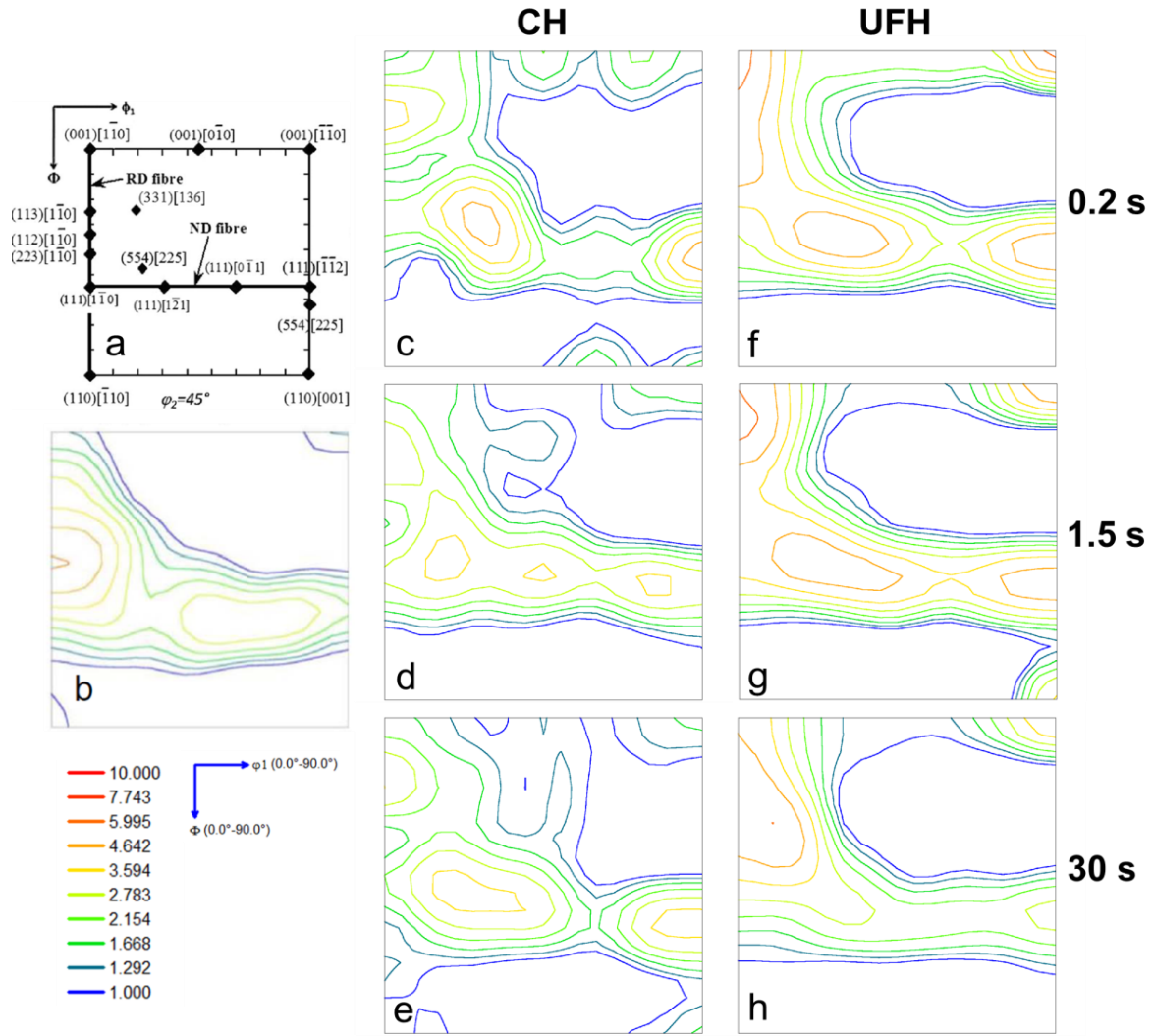

**Figure 10**: Effect of heating rate and soaking time on the orientation distribution function (ODF) of the studied material for $\varphi_2=45°$ in the Euler space; a) Ideal BCC texture components for $\varphi_2=45°$ in the Euler space; b) ODF of the initial cold rolled material, reproduced from [5]; c), d) and e) ODF corresponding to the CH conditions annealed for 0.2, 1.5 and 30 s, respectively; f), g) and h) correspond to the UFH conditions soaked for 0.2, 1.5 and 30 s, respectively.

*3.5. XRD analysis*

XRD measurements were carried out to analyze the evolution of retained austenite and its carbon content with soaking time. The results are listed in **Table 3** and compared to the values obtained by TKD (see Section 3.6).

Table 3: Effect of the heating rate and soaking time on the retained austenite volume fraction and its carbon content measured by XRD and TKD analysis.

| Condition | XRD (%) | % C (wt.) | TKD (%) |
|---|---|---|---|
| **CH10-0.2s** | 7.9 | 0.77 | 4.8 |
| **UFH800-0.2s** | 6.6 | 0.80 | 8.1 |
| **UFH800-1.5s** | 6.9 | 0.77 | 4.9 |
| **UFH800-30s** | 5.2 | 0.70 | 4.4 |

After short annealing (soaking for 0.2 s), the CH sample presents a higher retained austenite fraction compared to the UFH condition. The CH treatments lead to phase fractions closer to the ones at the equilibrium condition since there is more time for the austenite to nucleate and grow (**Table 2**). In the CH10-0.2s sample, taking into account fractions of both phases (i.e. retained austenite measured by XRD in **Table 3** and martensite determined by EBSD in **Table 2**), the total fraction of austenite formed during intercritical annealing is close to 20 %. The effect of soaking time on the retained austenite volume fraction for the UFH samples has two different trends. For short soaking times (0.2 s, 1.5 s), both nucleation and growth of intercritical austenite take place, as it is observed from the martensite fraction (see Section 3.3). Then, the volume fraction of austenite rises slightly from 6.6 % to 6.9 % with increasing time within the short range (**Table 3**). This effect indicates, that the nucleation stage plays a more important role compared to the growth stage, as there is a significant austenite fraction, which retains after rapid cooling, with a carbon concentration similar to the CH condition. Eventually, when the soaking time increases up to 30 s, the austenite fraction at the peak temperature increases due to the longer time to nucleate and grow, as there is a significant fraction of martensitic grains having a size below 1 μm (**Figure 9**), but its carbon concentration decreases up to 0.7 % reducing the amount of retained austenite down to 5.2 %.

The volume fractions of retained austenite measured by XRD (**Table 3**) are considerably higher than the values determined by EBSD (**Table 2**). This effect is produced by the large difference in the depth of the analyzed area being approximately 1 μm for XRD and 50 nm for EBSD [57]. As is well known, the metastable retained austenite generates a local increase in volume during transformation into martensite [58]. As phase transformation on the surface allows an easier accommodation of this volume change, the surface retained austenite grains are more prone to phase transformation during sample preparation, that reduces the amount of retained austenite detected by EBSD [57]. Meanwhile, XRD is able to detect retained austenite present in the bulk material, which has not transformed into martensite. Moreover, it should be noted that although the spatial resolution of the EBSD is reasonably high (65 nm in step size), it is not sufficient for detection of the finest austenite grains present in the microstructure, revealed by TEM analysis (see Section 3.6). Similar conclusions were drawn for other steel grades containing metastable austenite, such as Q&P steels in [59,60].

*3.6. TEM and TKD analysis*

To study the evolution of microstructure during soaking on nanoscale, TKD analysis combined with TEM characterization were carried out on CH10-0.2s and UFH after 0.2, 1.5 and 30 s samples. **Figure 11** represents the phase maps of the different samples analyzed by TKD. They are in a good accordance with the outcomes of the EBSD measurements presented above (see Section 3.3). Larger ferritic grains are observed in the CH10-0.2s samples (**Figure 11**a) compared to those seen in the UFH samples (**Figure 11**b, c, d). In addition, the CH treatment results in equiaxed ferritic grains without LAGBs in their interior (**Figure 11**a) due to the longer treatment time, while the UFH leads to an inhomogeneous microstructure with varying grain size and a higher fraction of LAGBs (**Figure 11**b, c, d).

Values of retained austenite volume fraction measured by TKD are provided in **Table 3**. They are higher compared to those determined by EBSD. This effect is caused by higher spatial resolution of the TKD technique, which enables to resolve nanoscale microstructural constituents having 10-30 nm in size [27]. Discrepancies between the volume fractions of retained austenite determined by XRD and those measured by TKD should also be noted. Unlike in the XRD measurements, a very local area is analyzed by TKD which leads to statistically insignificant data. Moreover, the TKD results highly depend on the quality of the studied samples. If the electropolishing step is inhomogeneous, there are significant differences in the foil thickness through the sample. If a local area is too thick, the electrons are unable to pass through and reach the detector, as their initial energy is orders of magnitude less compared to the ones generated in TEM which results in the non-indexed areas. Similar effect occurs when the foil is too thin, as too many electrons cross the specimen and reach the detector [26,61]. Diffraction patterns were taken from different austenitic regions observed by TKD in all samples, in order to prove the presence of austenite in the material, as it is shown in **Figure 11e)**.

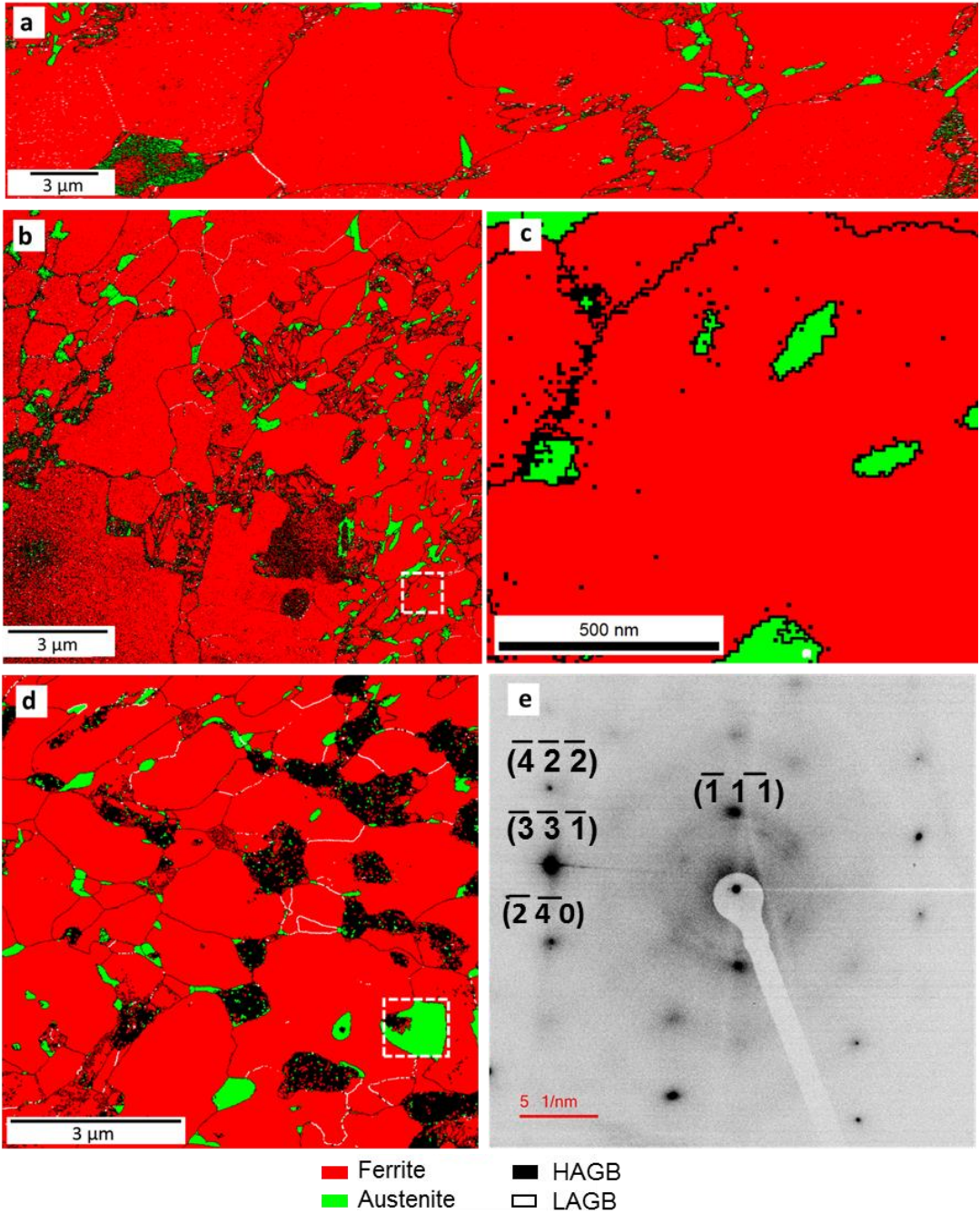

**Figure 11**: Phase maps obtained from TKD analysis in a) CH10-0.2s and UFH for 0.2 s (b & c), and 1.5 s (d)). Figure c) shows a detailed region in figure b). Figure e) represents the diffraction pattern of the austenite marked in figure d). Ferrite is shown in red and austenite in green. HAGBs are represented in black and LAGBs in white. Large regions in black are areas with a confidence index (CI) lower than 0.1.

**Figure 12**a, c, e shows TEM images illustrating microstructure evolution during UFH treatment of the steel within the non-recrystallized areas (as discussed in Sections above).

**Figure 12**b, d, f illustrate the corresponding KAM maps of the corresponding regions extracted from the TKD analysis. Formation of dislocation walls and other configurations is observed after UFH 0.2 s treatment, which are represented in form of lines with local misorientation < 1º on KAM maps (**Figure 12**a, b). Dislocation walls associated to recovery were reported elsewhere [49,62]. Longer soaking time of 1.5 s allows further dislocation climb and rearrangement and onset of LAGBs formation (**Figure 12**c, d). Finally, annealing for 30 s results in formation of an energetically favorable substructure in the grain interior (**Figure 12**e) with local misorientation at LAGBs reaching 4º (**Figure 12**f). In **Figure 12**e, f, enhanced local dislocation density and increased local misorientation are clearly seen also in the ferritic matrix near the martensite/ferrite interface (marked by white arrows). It is related to accommodation of the plastic micro-strain induced by the volume expansion due to the austenite/martensite transformation during rapid cooling. This observation was reported earlier for DP steels [63].

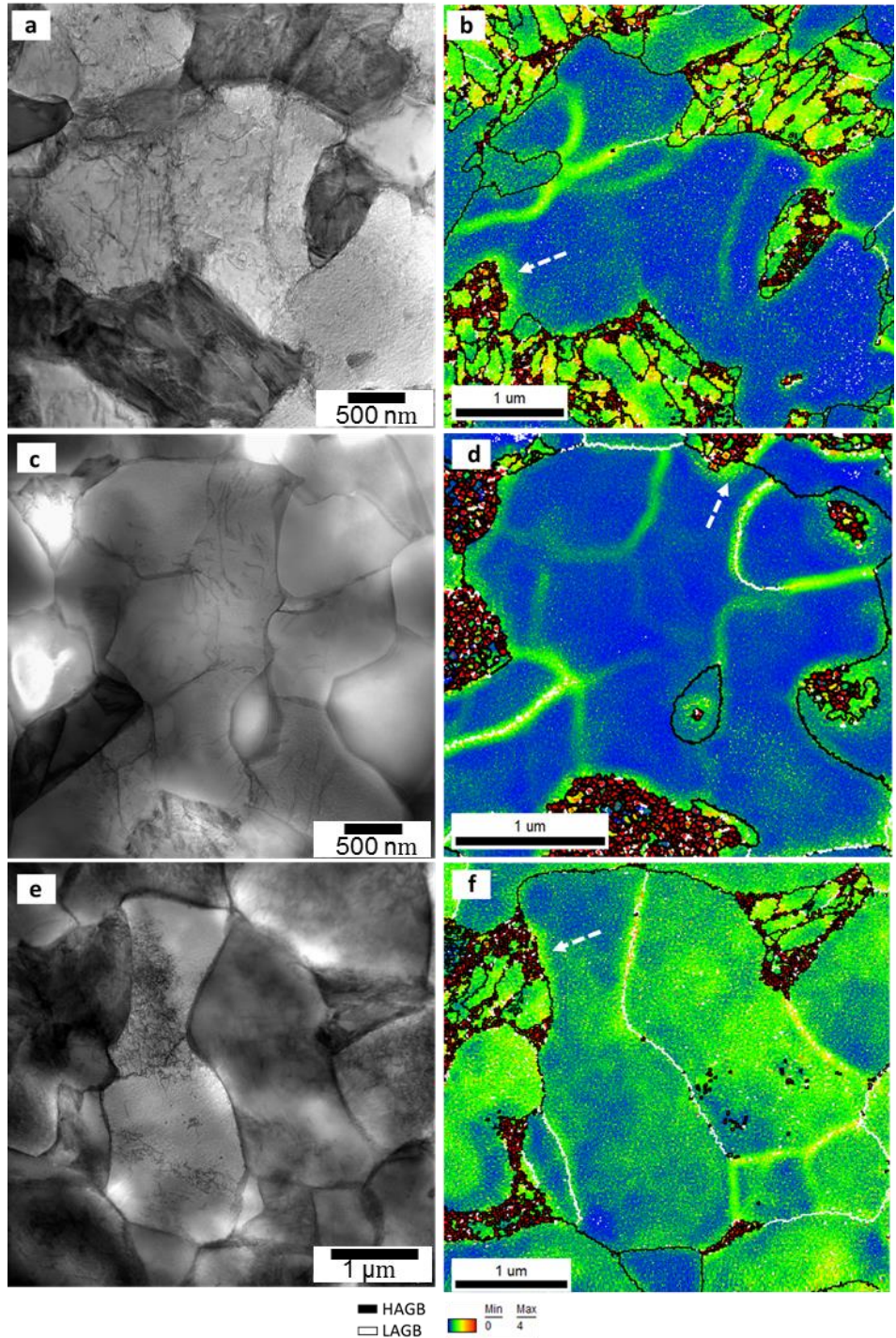

**Figure 12**: TEM images after UFH treatment for a) 0.2 s, c) 1.5 s and e) 30 s; KAM maps for b) 0.2 s, d) 1.5 s and f) 30 s obtained from the TKD analysis. White dashed arrows indicate the increase in misorientation in the ferritic matrix due to the martensite formation. (HAGBs in black, LAGBs in white).

The outcomes of this study clearly indicate that the microstructure of the low carbon steel is very sensitive to the soaking time at the peak temperature during UFH treatment. This provides an additional tool for microstructural design in carbon steels by manipulating also the soaking time in addition to the heating rate [5] and initial microstructure [12] of steels.

Grain size, volume fraction of martensite, volume fraction of non-recrystallized and recrystallized ferrite can be optimized via the correct balance of the heat treatment parameters, so steels with the excellent combination of high strength and ductility can be manufactured [5]. The approach can be applied to all carbon steels.

## 4. Conclusions

The effect of heating rate and soaking time on the microstructure of the heat-treated low carbon steel was studied using SEM, EBSD, XRD, TKD and TEM techniques. The following conclusions can be drawn.

1. A complex multiphase, hierarchic microstructure mainly consisting of ferritic matrix with embedded martensite and retained austenite is formed after all applied heat treatments. There is significant effect of soaking time on the microstructure of the UFH treated steel, while it does not affect the microstructure evolved in the CH treated material.
2. There is a strong effect of heating rate on the microstructure of the ferritic matrix. The CH treatment results in the ferritic matrix consisting mainly of equiaxed recrystallized grains independently on the soaking time, while fine recrystallized grains and larger non-recrystallized (i.e. recovered) ferritic grains are present in all UFH treated conditions. The fraction of recrystallized ferritic grains generally tends to increase with increasing soaking time. Combined TEM and TKD study proved directly that the recovery process starts with formation of dislocation walls via dislocation climb and rearrangement, which gradually transform into LAGBs.
3. Volume fraction of martensite tends to increase with increasing soaking time during UFH treatment due to suppression of cementite spheroidization, which, in turn, reduces the amount of energetically favorable sites for austenite nucleation and results in longer soaking time to reach the equilibrium at the inter-critical peak temperature.
4. Based on the outcomes of the XRD analysis, it is possible to conclude that UFH treatments results in slightly lower amount of retained austenite compared to CH treatment. The amount of retained austenite and carbon content therein tend to slightly decrease with increasing soaking time after UFH treatment due to lower carbon gradients in the material before rapid cooling.

5. TKD analysis allows to precisely identify and analyze the retained austenite nanograins and other nanoscale elements of the complex microstructure along with the local misorientations due to dislocation generation and rearrangement.
6. TKD and TEM proved that local volume expansion due to austenite-martensite phase transformation during rapid cooling induces dislocations into the ferritic grains.

**Acknowledgements**

MAVT acknowledges gratefully the financial support by IMDEA Innovation Award.

**Data availability statement**

The raw/processed data required to reproduce these findings cannot be shared at this time as the data also forms part of an ongoing study.